\documentclass[twocolumn,prl,floatfix,superscriptaddress]{revtex4}
\usepackage{graphics}
\usepackage{graphicx}
\usepackage{dcolumn} 
\usepackage{bm}
\usepackage{epsfig}
\usepackage{amsmath}
\usepackage{amsfonts}
\usepackage{amssymb}
\usepackage{times}
\newcommand{\eg}{{\sl e.g.}}

\begin{document}

\title{Parallel electron-hole bilayer conductivity from electronic interface reconstruction}

\author{R. Pentcheva}
\affiliation{Department of Earth and Environmental Sciences, University of Munich, Germany}
\author{M. Huijben}
\affiliation{Faculty of Science and Technology and MESA$^+$ Institute for Nanotechnology, University of Twente, The Netherlands}
\author{K. Otte}
\affiliation{Department of Earth and Environmental Sciences, University of Munich, Germany}
\author{W.E. Pickett}
\affiliation{Department of Physics, University of California, Davis, USA}
\author{J.E. Kleibeuker}
\affiliation{Faculty of Science and Technology and MESA$^+$ Institute for Nanotechnology, University of Twente, The Netherlands}
\author{J. Huijben}
\affiliation{Faculty of Science and Technology and MESA$^+$ Institute for Nanotechnology, University of Twente, The Netherlands}
\author{H. Boschker}
\affiliation{Faculty of Science and Technology and MESA$^+$ Institute for Nanotechnology, University of Twente, The Netherlands}
\author{D. Kockmann}
\affiliation{Faculty of Science and Technology and MESA$^+$ Institute for Nanotechnology, University of Twente, The Netherlands}
\author{W. Siemons}
\affiliation{Department of Physics, University of California, Berkeley, USA}
\author{G. Koster}
\affiliation{Faculty of Science and Technology and MESA$^+$ Institute for Nanotechnology, University of Twente, The Netherlands}
\author{H.J.W. Zandvliet}
\affiliation{Faculty of Science and Technology and MESA$^+$ Institute for Nanotechnology, University of Twente, The Netherlands}
\author{G. Rijnders}
\affiliation{Faculty of Science and Technology and MESA$^+$ Institute for Nanotechnology, University of Twente, The Netherlands}
\author{D.H.A. Blank}
\affiliation{Faculty of Science and Technology and MESA$^+$ Institute for Nanotechnology, University of Twente, The Netherlands}
\author{H. Hilgenkamp}
\affiliation{Faculty of Science and Technology and MESA$^+$ Institute for Nanotechnology, University of Twente, The Netherlands}
\author{A. Brinkman}
\affiliation{Faculty of Science and Technology and MESA$^+$ Institute for Nanotechnology, University of Twente, The Netherlands}

\date{\today}

\begin{abstract}
The perovskite SrTiO$_3$-LaAlO$_3$ structure has advanced to a model system  to investigate the rich electronic phenomena arising at polar interfaces.
Using first principles calculations and transport measurements we demonstrate that an additional SrTiO$_3$ capping layer prevents atomic reconstruction at the LaAlO$_3$ surface and  triggers the electronic reconstruction at a significantly lower LaAlO$_3$ film thickness than for the uncapped systems.  Combined theoretical and experimental evidence (from magnetotransport and ultraviolet photoelectron spectroscopy) suggests two spatially separated sheets with electron and hole carriers, that are as close as 1 nm. 
\end{abstract}

\maketitle

Polarity discontinuities at the interfaces between different crystalline materials are usually compensated by atomic reconstructions via defects or adsorbates as in conventional 
semiconductor interfaces. However, in complex oxides the mixed valence states provide an extra option for charge rearrangement by redistributing electrons at lower energy cost than redistributing ions. The remarkable electronic transport properties that occur at the interface between the band insulators SrTiO$_3$ (STO) and
LaAlO$_3$ (LAO)  \cite{ohtomo2004,ohtomo2006,nakagawa2006,huijben2006,thiel2006,reyren2007,brinkman2007,caviglia2008,basletic2008} have been attributed to this so called electronic reconstruction\cite{Sawatzky,millis2004} but direct evidence has not yet been found.

The polarity of LAO arises from the LaO and AlO$_2$ layers being not charge neutral in the [001] direction, unlike the formally neutral TiO$_2$ and SrO layers of STO. In the ionic limit, LaO has a charge $q=$ +e and AlO$_2$ $q=-$e per unit cell. The screened dipole per unit cell is then $D=q \Delta z /\epsilon$, where the spacing $\Delta z = c/2$ ($c=3.9$~\AA\ is the out of plane lattice parameter) and $\epsilon$ = 25 is the dielectric constant of LAO \cite{hayward2005}. Screening contributions come primarily from a strong lattice polarization of the LAO film (their contribution can be as high as $\sim 62\%$ \cite{pentcheva2009}), supplemented by electronic cloud deformation \cite{ishibashi2008}. For STO-LAO systems, the remaining screened dipole of 0.08 e\AA~per cell is expected to give rise to an internal electric field of $2.4 \times 10^7$ V/cm, and a resulting build-up of electric potential of 0.9 V per LAO unit cell. 

This potential shift explains quantitatively why, above a threshold of 3-4 unit cells, electrons are transferred from the surface, across the LAO slab, into the STO conduction band. The resulting insulator-to-metal transition has been observed experimentally for the $n$-type LaO/TiO$_2$-interface \cite{thiel2006}. However, the corresponding potential shifts across LAO have not been detected so far in experiments, which suggests that possibly non-electronic reconstructions occur during the growth, driven by the polar potential build-up. For a reconstructed STO-LAO interface, it should be noted that to avoid potential build-up also the LAO surface itself needs to reconstruct, either structurally, electronically, or chemically. After electronic charge transfer one would expect holes at the surface, which have also never been observed. 

In this Letter we show that an additional STO capping layer circumvents structural and chemical reconstructions at the LAO surface. The O $2p$ band in the STO capping layer allows for hole doping, so that an electronic reconstruction mechanism comes into play. By means of the STO capping layer one enters a new regime in the field of electronically reconstructed oxide interfaces with two spatially separated 2D conducting sheets, one electron-like and the other hole-like, that can display new electronic behavior including the possibility of a 2D excitonic liquid phase.

\begin{figure}
     \begin{center}
\includegraphics[scale=0.95]{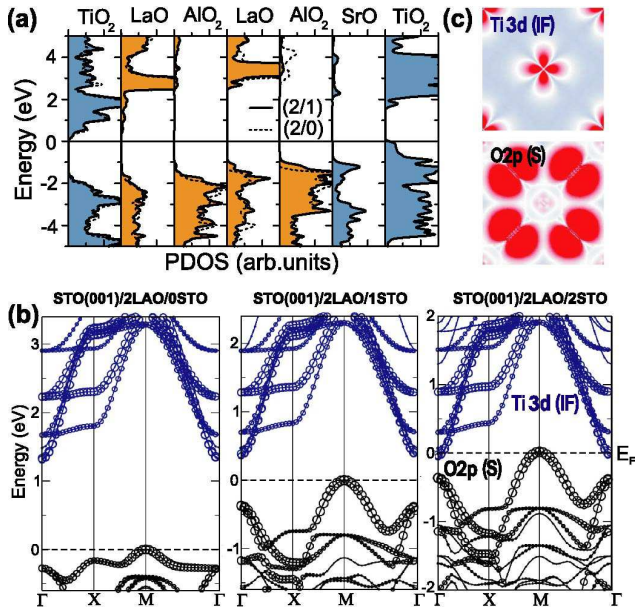}
    \end{center}
\caption{\label{fig:bandshifts} (a) Layer-resolved density of states (DOS) of STO(001)/2LAO (dotted line), and STO(001)/2LAO/1STO (black line, colored area) aligned at the bottom of the Ti $3d$ band at the interface. 
(b) Influence of the STO capping on the band structure of STO(001)/2LAO/$m$STO with increasing number of capping layers ($m=0-2$), showing the closing of the band gap due to overlap between surface O $2p$ states (black circles) and interface Ti $3d$ states (blue circles). 
(c) The electron density distribution in the TiO$_2$ layers of the $m=2$ sample shows at the interface electrons in the Ti $3d_{xy}$ orbitals (top) and holes in the O $2p_{\pi}$ bands at the surface (bottom). The electron density is integrated between -0.6 and 0.0~eV.}
\end{figure}

The system consisting of a varying number of LAO monolayers (ML), $n=1-5$~ML, and of a STO capping layer, $m=0-2$~ML, stacked on an STO(001) substrate, was studied by DFT calculations in the generalized gradient approximation (GGA) (for details on the calculations see Refs. \cite{pentcheva2009,EPAPS}). The calculated layer-resolved densities of states are presented in Fig.~\ref{fig:bandshifts}a for 2ML LAO with and without 1ML STO capping. The effect of the electric field within the LAO film is apparent from the shifts of bands,  \eg\ by $\sim$0.4 eV per LAO unit cell for the uncapped system \cite{pentcheva2009}. Note, that this potential shift is smaller than the mentioned 0.9 eV due to effects related to the well-known underestimation of band gaps by density functional theory (DFT).

Adding a single unit-cell STO capping layer is found to have a dramatic impact on the calculated electronic structure: the band gap, being 1.2~eV for STO(001)/2LAO, is nearly closed for STO(001)/2LAO/1STO. While the ionic relaxation pattern \cite{pentcheva2009} changes significantly when a capping layer is added \cite{EPAPS}, the net contribution of the buckled TiO$_2$ and SrO layers does not affect appreciably the total ionic dipole moment of the film (which scales with the number of LAO layers). Hence, the gap reduction  has mainly an electronic origin. 

The evolution of the band structure of STO(001)/2LAO/$m$STO with increasing number of capping layers ($m=0-2$) is depicted in Fig.~\ref{fig:bandshifts}b. The valence band maximum is defined by the O $2p$-states at the M($\pi$,$\pi$)-point in the surface layer, while Ti $3d$-states at $\Gamma$ at the $n$-type interface mark the bottom of the conduction band.  In the capped systems a dispersive O $2p$ surface band extends 0.8 eV above the subsurface O $2p$ band and effectively reduces the band gap driving the insulator-to-metal transition at an LAO thickness of only 2ML compared to 4 ML in the uncapped case. This surface state is analogous to the one on the clean STO (001) surface~\cite{kimura1995,padilla1998}. Further STO layers (\eg\ STO(001)/2LAO/2STO \cite{EPAPS}) increase the band overlap at the Fermi level, but have an overall weaker influence due to the lack of internal field in STO.
The reduction of the band gap, and finally its closing, is thus due to three \textit{electronic} effects: (i) the steady upward shift of the O $2p$ states as they approach the surface \cite{pentcheva2009} due to the internal polarity of LAO, (ii) the band discontinuity at the interface between LAO and the capping STO layer, and (iii) the dispersive O $2p$ surface band in the capped systems that extends 0.8 eV above the subsurface O $2p$ band.  

\begin{figure} %[!h]
    \vspace{-3pt} 
     \begin{center}
\includegraphics[scale=0.90]{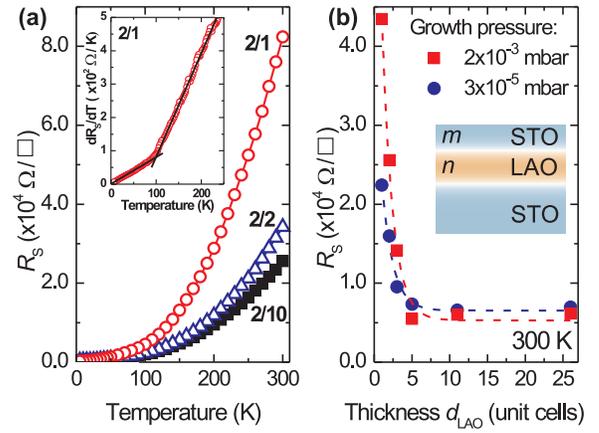}
    \end{center}
\caption{\label{fig:transportdata} 
(a) Sheet resistance as function of temperature for three different STO-LAO-STO samples: STO(001)/2LAO/1STO (red circles), STO(001)/2LAO/2STO (blue triangles), and STO(001)/2LAO/10STO (black squares). The sample of STO(001)/2LAO was found to be insulating. All samples are grown at $2 \times 10^{-3}$ mbar of oxygen. Inset: $dR/dT$ as function of temperature with different linear fits below and above 100 K. 
(b) Sheet resistance at room temperature of STO(001)/$n$LAO/10STO samples for varying $n$ LAO interlayer thickness and a fixed number of 10 unit cells of STO capping layer. Red squares indicate samples grown at relatively high oxygen pressure ($2 \times 10^{-3}$ mbar), blue circles indicate samples grown at lower oxygen pressure ($3 \times 10^{-5}$ mbar).}
    \vspace{-10pt} 
\end{figure}

Experimentally, we confirm the crucial influence of a single monolayer of nonpolar material on the electronic interface reconstruction. STO(001)/$n$LAO/$m$STO samples were made by pulsed laser deposition of $n$ ML of LAO and $m$ ML of STO on TiO$_2$-terminated STO(001) substrates (for fabrication details, see Ref.~\onlinecite{huijben2006,brinkman2007}). While uncapped STO(001)/2LAO samples are found to be insulating (sheet resistance above 1 G$\Omega/\square$), samples with an additional single ML of STO are conducting (see Fig.~\ref{fig:transportdata}a). The conductivity is further enhanced in STO(001)/2LAO/2STO samples, but the influence of increasing the STO capping layer thickness weakens, as expected from the DFT results: the STO(001)/2LAO/10STO sample has almost the same conductivity as the STO(001)/2LAO/2STO sample. Samples with a single ML of LAO were found to be insulating except for the sample with a thick STO capping ($m\geq10$). 

It is known that the sheet resistance in STO/LAO samples depends critically on the oxygen pressure during growth~\cite{brinkman2007,siemons2007,herranz2007,kalabukhov2007,huijben2009} and can vary over many orders of magnitude \cite{brinkman2007}.  
Figure~\ref{fig:transportdata}b shows the sheet resistance for two different sets of STO/LAO/STO heterostructures with varying LAO interlayer thickness, grown at a relatively high oxygen pressure ($2 \times 10^{-3}$ mbar) and at lower oxygen pressure ($3 \times 10^{-5}$ mbar). 
For the coupled-interface samples, the influence of the oxygen pressure is now found to be much weaker. Apparently, the STO capping protects the underlying LAO surface against reconstruction via defects or adsorbates.  

\begin{figure} %[!h]
     \begin{center}
\includegraphics[scale=0.95]{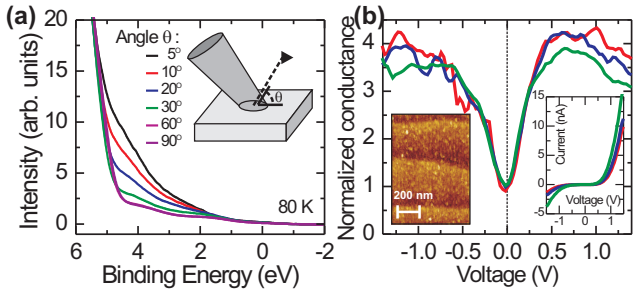}
    \end{center}
\caption{\label{fig:UPS} 
(a) UPS spectra of a STO(001)/2LAO/1STO sample taken \textit{in situ} after growth at 80 K under various detector angles (the inset shows how the angle towards the detector is defined). A gradual filling of the valence band towards the Fermi energy is shown for lower angles. Lower angle spectra have larger contributions from the surface layers.
(b) STS conductance $(dI/dV)/(I/V)$ at 300 K for different tip-sample distances (current set-point respectively 1.5 nA, 2.0 nA, and 4.0 nA at a bias voltage of -1.5 V). The Fermi energy, $E_F$, at $V=0$~V (blue dashed line) is found to lie between the valence band at the M-point of the STO capping surface and the conduction band at the substrate-LAO interface. The right inset shows the current-voltage characteristics from which the conductance was derived and the left inset shows an STM topography image taken at 300 K with a bias voltage of -1.0 V, revealing the substrate steps.}
\end{figure}

In order to obtain spectroscopic evidence for the \textit{electronic} reconstruction, ultraviolet photoelectron spectroscopy (UPS) was performed \textit{in situ} immediately after the growth of a STO(001)/2LAO/1STO sample. 
Figure \ref{fig:UPS}a shows a gradual increase in intensity for the more surface sensitive measurements at lower detector angles. These states originate from the valence band of LAO as well as the valence band of the STO surface ML.
Note, that the valence band states penetrate all the way to $E_F$, unlike studies on doped STO \cite{UPS}, where only trapped states close to the conduction band are usually observed. 

To probe states around the Fermi level, scanning tunneling spectroscopy (STS) was performed in ultra high vacuum using a variable temperature cryostat. Figure \ref{fig:UPS}b shows the local density of states, $(dI/dV)/(I/V)$, of a STO(001)/2LAO/1STO sample. At room temperature, the Fermi energy lies between the valence band of the STO surface and the conduction band of the substrate-LAO interface. The bandgap is (almost) closed, as predicted in Fig.~\ref{fig:bandshifts} by DFT (for more spectroscopic details, see \cite{EPAPS}). The observed density of states just below the Fermi energy, as measured by both UPS and STS, is consistent with the electronic reconstruction scenario and suggests the presence of holes in transport.

\begin{figure}%[!h]
     \begin{center}
\includegraphics[scale=0.95]{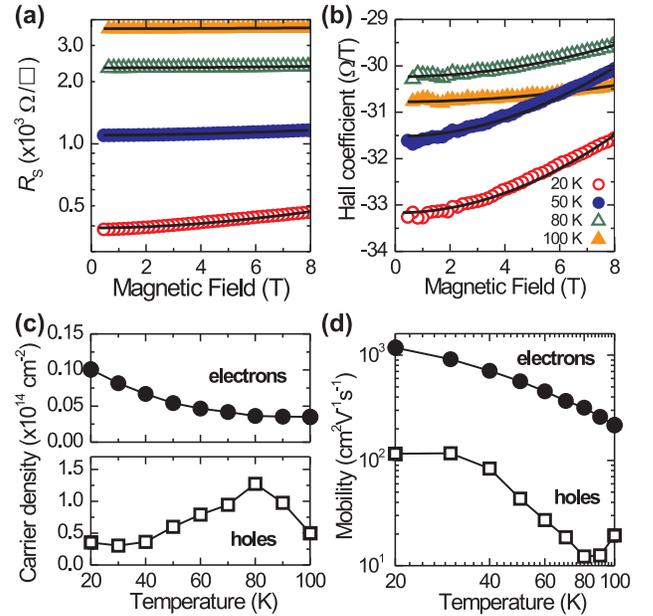}
    \end{center}
\caption{\label{fig:MRexp} 
(a) Sheet resistance as function of magnetic field at different temperatures for a STO(001)/2LAO/1STO sample, exhibiting a positive magnetoresistance. (b) Hall coefficient ($R_H/H$) of the same sample as a function of magnetic field. (c) Sheet carrier densities and (d) mobilities as obtained from a two-band fit to the magnetoresistance and Hall data at each temperature.}
\end{figure}

To investigate the possibility of a parallel electron-hole bilayer and the sign of the charge carriers in capped systems, magnetoresistance and Hall data were analyzed. Because the intrinsic coupling between the layers would not allow to probe the transport properties of the layers individually, unless structures are realized on a sub-micron length scale, our measurements contain information on the layers in parallel. Fig.~\ref{fig:MRexp} displays  a positive non-quadratic magnetoresistance and a Hall resistance whose slope increases for higher fields for all conducting STO/LAO/STO samples. Quantum oscillations can still be excluded because of the low mobility. A negative magnetoresistance contribution, observed for single-interface samples deposited at high oxygen pressure \cite{brinkman2007}, only appears below 10 K. It is natural to interpret the observations in terms of multiband conductivity. Indeed, in the temperature range up to 100~K, the magnetoresistance as well as the Hall resistance can be fitted with a two band model \cite{EPAPS} (solid lines in Fig.~\ref{fig:MRexp} a and b). Two carrier concentrations and two mobilities could be obtained for the STO(001)/2LAO/1STO sample from fitting as a function of temperature (Fig.~\ref{fig:MRexp} c and d).   

The positive carrier sign of one of the bands at low temperatures indicates hole-type conductivity, while the other band is of electron-type. We note that no fit to the data could be obtained for equal signs of the two carrier densities. Neither oxygen vacancy doping, nor doping by cation substitution, have ever been shown to give rise to hole conductivity in the STO/LAO system. The calculated electron density distribution in Fig.~\ref{fig:bandshifts}c displays electrons of Ti $3d_{xy}$ orbital character in the interface TiO$_2$ layer, while holes of O $2p_{\pi}$ type are present in the surface TiO$_2$ layer. Consequently, we attribute the hole band to the surface layer, while the  electron band, with a lower carrier density but a much larger mobility, is naturally attributed to the Ti $3d_{xy}$ states at the interface to the substrate, consistent with the observations in Ref. \cite{salluzzo}. 

Note, that the hole density is about an order of magnitude larger than the electron density. However, the Hall effect is dominated by the electron band because of its large mobility (10$^{3}$ cm$^{2}$V$^{-1}$s$^{-1}$, similar to  values obtained on single interface STO/LAO samples deposited at oxygen pressures $>10^{-3}$ mbar). The unequal number of electrons and holes illustrates that not all charge carriers are visible in transport measurements. While the effective electron and hole masses cannot be directly inferred from our data, the band structure calculations (neglecting strong correlation effects and defects) render 0.4 $m_e$ for the electrons (both in the capped and uncapped system) and a significantly higher effective mass of 1.2 $m_e$ for the holes. Above 100 K, neither magnetoresistance, nor a nonlinear Hall resistance were observed, because the mobilities become so low that no magnetoresistance effects are expected any longer ($\mu^2H^2\ll1$ in the two-band equations of \cite{EPAPS}). Therefore, no two-band fitting analysis can be performed in this case, and no statement can be done on the presence of electrons and holes.

In the STO(001)/2LAO/1STO sample evidence from different experimental techniques point to an electronic reconstruction mechanism. At the same time, we know that for uncapped thick LAO samples no potential build up has been observed \cite{segal2009}, suggestive of non-electronic reconstruction scenarios. In order to verify whether it is only the capping that makes a difference, the magnetotransport was studied for a large number of samples with either thicker LAO or thicker capping STO \cite{EPAPS}. While  for defect-free systems theory predicts an increase in the band overlap, and hence in the number of electrons and holes, no hole contribution was found experimentally beyond $n=2$ and $m=1$. Apparently, during growth structural reconstruction occurs whenever the potential build up exceeds a few eV. While conductivity arises also in this case, no evidence for pure electronic reconstruction exists any longer. The STO(001)/2LAO/1STO sample might be an example of a structure where the potential build-up during the growth of just 2 ML of LAO is just not yet large enough for a reconstruction. The subsequent capping ML protects the LAO surface from structural or chemical reconstruction, provides another potential increase (either during growth or cooldown), and can accomodate mobile holes, resulting in electronic reconstruction. This conclusion provides guidelines to enhancing electronic reconstruction effects in general.

As shown in Fig.~\ref{fig:bandshifts}, the surface valence band has its maximum at the M=($\pi,\pi$) zone corner, whereas the substrate-LAO interface conduction band has its minimum at the zone center. This makes the band overlap distant not only in real space (across 12~\AA~or more, depending on capping layer thickness) but also indirect in momentum. As a practical consequence, an electron at the surface cannot move to the substrate without some mechanism to supply the momentum transfer. The obvious mechanism is via phonons, specifically M=($\pi,\pi$) phonons. These are zone boundary optical phonons, which typically lie at a few tens of meV energy. Equilibration of electrons and holes across the LAO slab will be slow at low temperature, but will occur rapidly as soon as optical phonons are excited. 

In summary, the STO capping has enabled us to show that holes are present in electronically reconstructed oxide interface samples. Their mobility is low and it is expected that the holes can become localized or eliminated in uncapped STO/LAO systems more strongly (\eg\ by absorbed molecules or by ionic surface reconstruction). This possibly explains the large sensitivity of uncapped samples to growth conditions and the possibility to manipulate the interface conductivity by an atomic force microscope tip \cite{cen2008, cen2009}.

A further consequence of this 2D electron-hole bilayer is that it provides the conditions necessary for formation of a 2D excitonic liquid \cite{keldysh1965,cloizeaux1965} comprised of interacting indirect excitons. In this oxide nanostructure the separation of the 2D electron and hole gases can be varied by the choice of polar material as well as capping material. In analogy to other oxides, such as ZnO \cite{Kawasaki}, it is expected that higher mobilities can be obtained by reducing the defect density. Furthermore, the carrier densities can be tuned by gating, allowing a substantial parameter range to be probed. 

This work is financially supported by the Dutch Foundation for Fundamental Research on Matter (FOM), the Netherlands Organization for Scientific Research (NWO) through VIDI and VICI grants, NANONED, the Bavaria-California Technology Center (BaCaTeC), German Science Foundation (TRR80) and DOE's Computational Materials Science Network, DOE Grant DE-FG02-04ER46111, and a grant for computational time at the supercomputer HLRBII at the Leibniz Rechenzentrum.

\end{document}

\title{Supplementary information for "Parallel electron-hole bilayer conductivity from electronic interface reconstruction"}

\author{R. Pentcheva}
\affiliation{Department of Earth and Environmental Sciences, University of Munich, Germany}
\author{M. Huijben}
\affiliation{Faculty of Science and Technology and MESA$^+$ Institute for Nanotechnology, University of Twente, The Netherlands}
\author{K. Otte}
\affiliation{Department of Earth and Environmental Sciences, University of Munich, Germany}
\author{W.E. Pickett}
\affiliation{Department of Physics, University of California, Davis, USA}
\author{J.E. Kleibeuker}
\affiliation{Faculty of Science and Technology and MESA$^+$ Institute for Nanotechnology, University of Twente, The Netherlands}
\author{J. Huijben}
\affiliation{Faculty of Science and Technology and MESA$^+$ Institute for Nanotechnology, University of Twente, The Netherlands}
\author{H. Boschker}
\affiliation{Faculty of Science and Technology and MESA$^+$ Institute for Nanotechnology, University of Twente, The Netherlands}
\author{D. Kockmann}
\affiliation{Faculty of Science and Technology and MESA$^+$ Institute for Nanotechnology, University of Twente, The Netherlands}
\author{W. Siemons}
\affiliation{Department of Physics, University of California, Berkeley, USA}
\author{G. Koster}
\affiliation{Faculty of Science and Technology and MESA$^+$ Institute for Nanotechnology, University of Twente, The Netherlands}
\author{H.J.W. Zandvliet}
\affiliation{Faculty of Science and Technology and MESA$^+$ Institute for Nanotechnology, University of Twente, The Netherlands}
\author{G. Rijnders}
\affiliation{Faculty of Science and Technology and MESA$^+$ Institute for Nanotechnology, University of Twente, The Netherlands}
\author{D.H.A. Blank}
\affiliation{Faculty of Science and Technology and MESA$^+$ Institute for Nanotechnology, University of Twente, The Netherlands}
\author{H. Hilgenkamp}
\affiliation{Faculty of Science and Technology and MESA$^+$ Institute for Nanotechnology, University of Twente, The Netherlands}
\author{A. Brinkman}
\affiliation{Faculty of Science and Technology and MESA$^+$ Institute for Nanotechnology, University of Twente, The Netherlands}

\maketitle

%-----------------------------------------------------------------------------
\section{Evolution of density of states}

Figure 1 shows the  evolution of the total density of states (DOS) as a function of the number, $m$, of STO capping layers in STO(001)/2LAO/$m$STO samples. While the first layer has the dominating effect of reducing the band gap by 1.2 eV, a second STO layer leads to a  clear band overlap and an enhancement of the DOS at the Fermi level.

% FIG S1
\begin{figure}[!h]
    \vspace{-10pt} 
     \begin{center}
\includegraphics[scale=0.7]{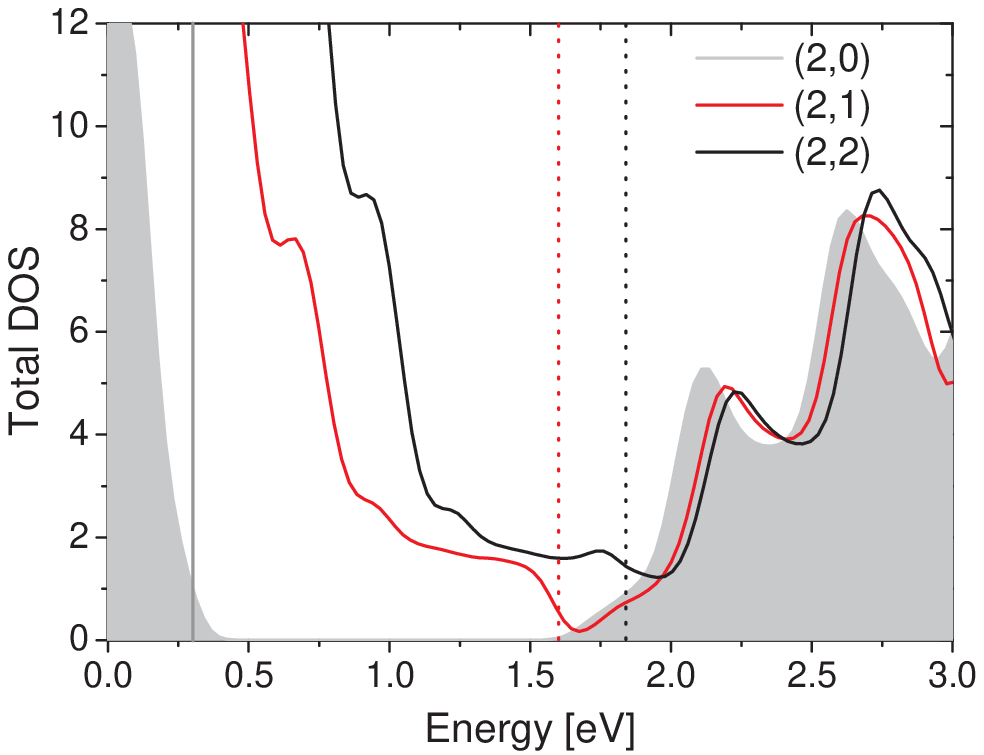}
    \end{center}
\caption{\label{fig:tdos} Total density of states (DOS) of STO(001)/2LAO/$m$STO, aligned at the bottom of the Ti $3d$ band at the interface. Vertical lines mark the positions of the Fermi level for each system. By adding a STO capping layer, the band gap of STO(001)/2LAO is reduced by 1.2 eV. Further capping layers lead to an increase of DOS at the Fermi level. }
    \vspace{-10pt} 
\end{figure}

%-----------------------------------------------------------------------------
\section{Ionic relaxations}

\begin{figure}[!h]
\vspace{-10pt} 
     \begin{center}
\includegraphics[scale=0.29]{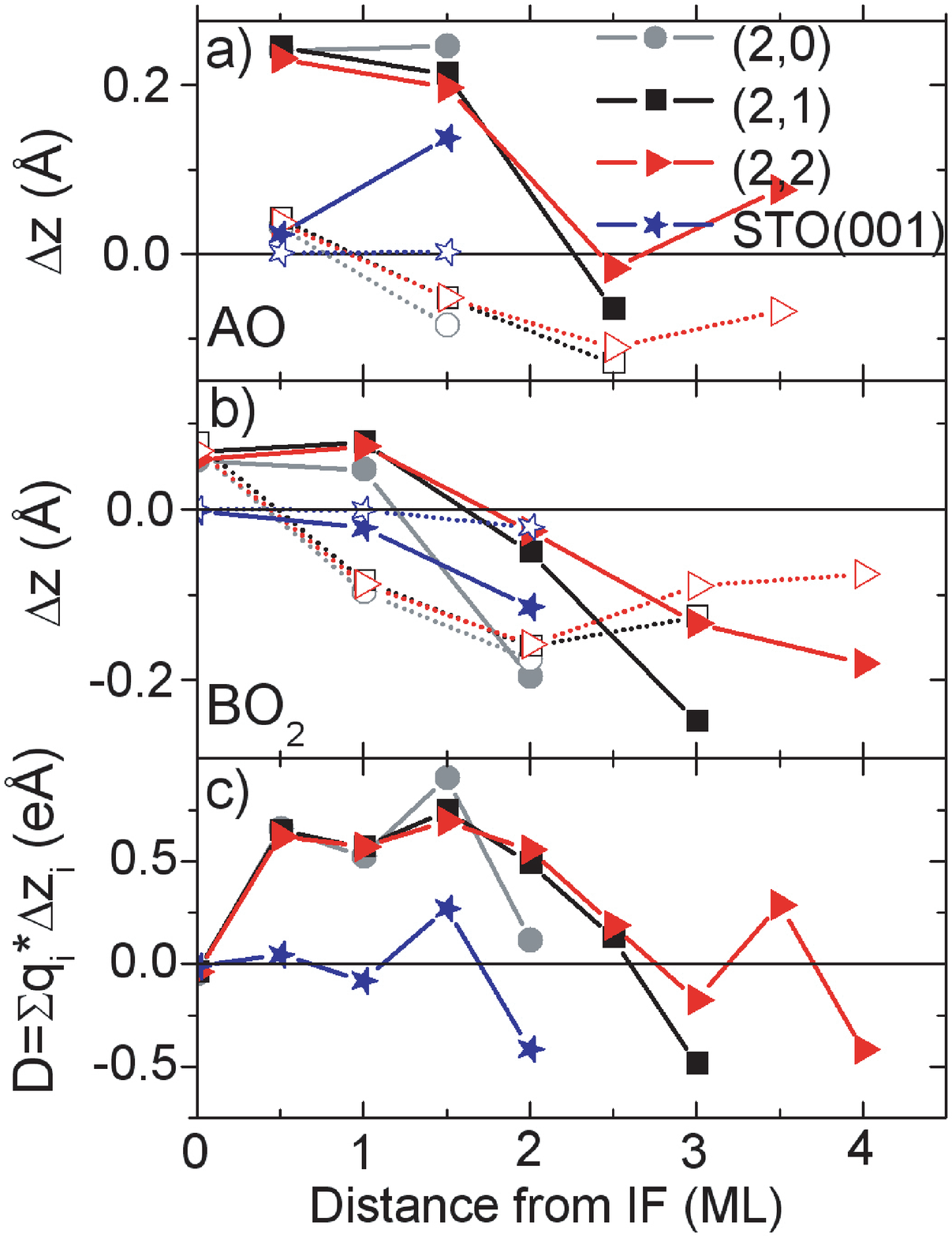}
    \end{center}
\caption{Ionic relaxations in STO(001)/2LAO/$m$STO and the bare STO(001)-surface. Vertical displacements $\Delta z$ are shown with respect to the bulk position in the (a) AO and (b) BO$_2$ layers, (c) layer-resolved dipole moments as a function of the distance from the interface (IF) TiO$_2$-layer. The filled (open) symbols mark cation (anion) relaxations. $(n,m)$ denotes the number $n$ of LAO and $m$ of STO layers in the respective system.  }
    \vspace{-10pt} 
\end{figure}

Figure 2 shows the calculated layer resolved ionic displacements~\cite{footnote1} in STO(001)/2LAO/$m$STO. Additionally, the relaxations of a STO(001)-surface are plotted. For the uncapped systems Al and O both relax inwards by similar amounts without a ferroelectric distortion. In contrast, there is a strong buckling in the surface TiO$_2$ layer in the case of STO capping. The  relaxation pattern in the capping layers bears a striking resemblance to the structure of the STO(001) surface, where the total dipole is relatively small $D_{ionic}^{\rm STO(001)}=-0.19$ e\AA~\cite{footnote2}. As mentioned previously, also the electronic structure of the STO(001) surface~\cite{kimura1995,padilla1998} and the capping layer are similar, in particular a dispersive O 2$p$ surface state with maximum at the M-point appears. Due to the small ionic contribution of the capping layer the total dipole moment is not affected appreciably by the capping layer: $D_{ionic}^{(2,0)}=2.15$ e\AA, $D_{ionic}^{(2,1)}=2.05$ e\AA, and $D_{ionic}^{(2,2)}=2.28$ e\AA. The latter  turns out to be determined by the total number of LAO layers, \eg\ $D_{ionic}^{(1,1)}=1.02$ e\AA, $D_{ionic}^{(2,1)}=2.05$ e\AA, and $D_{ionic}^{(3,1)}=3.30$ e\AA.

\section{Spectroscopic details}

\begin{figure}%[!h]
    \vspace{-10pt} 
     \begin{center}
\includegraphics[scale=0.62]{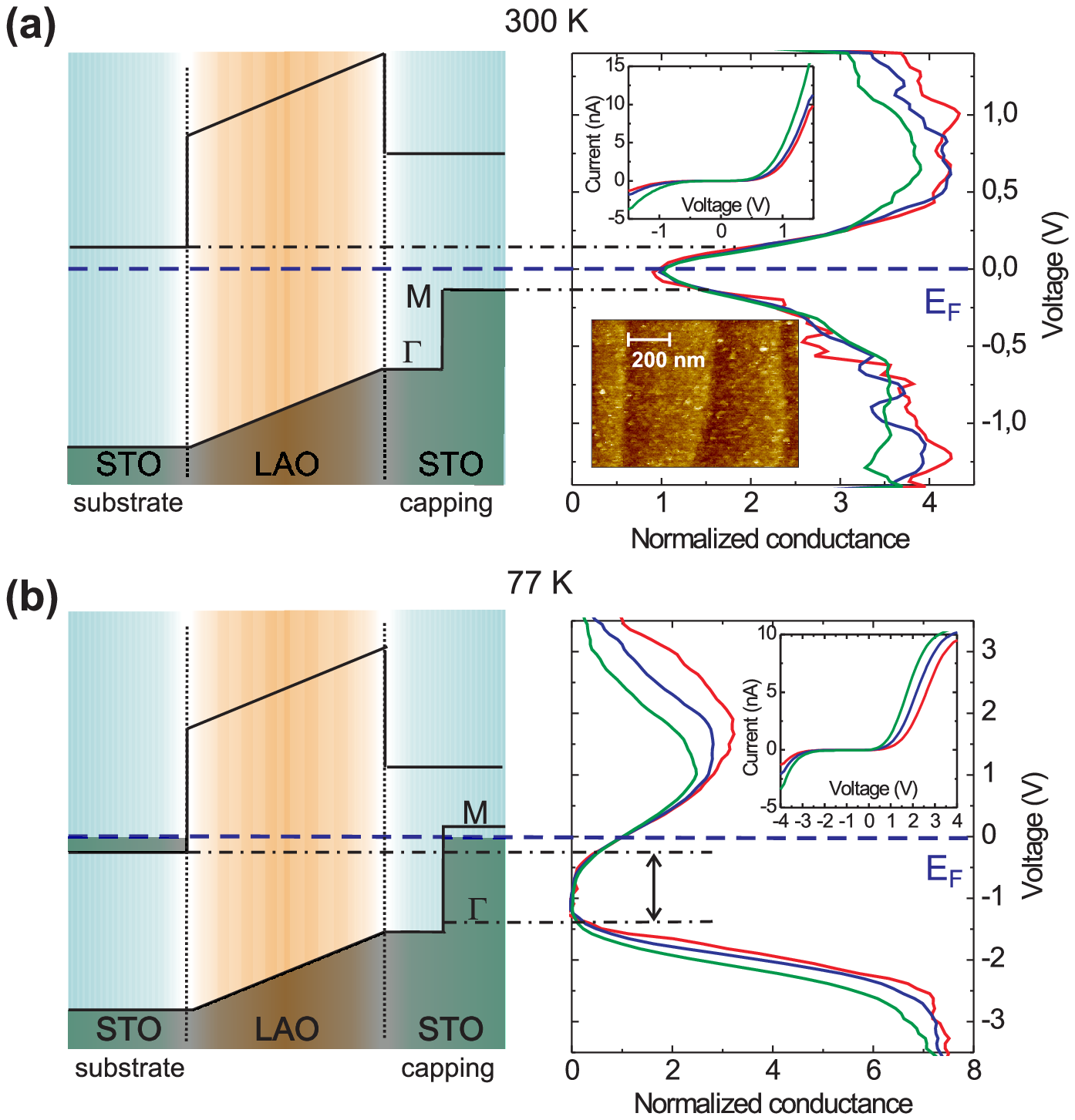}
    \end{center}
\caption{\label{fig:STM} (a) Scanning tunneling normalized conductance $(dI/dV)/(I/V)$ measurements of a STO(001)/2LAO/1STO sample at 300 K for different tip-sample distances (current set-point respectively 1.5 nA, 2.0 nA, and 4.0 nA at a bias voltage of -1.5 V). The schematic drawing explains the small gap between the valence band at the M-point of the STO capping surface and the conduction band at the substrate-LAO interface. The Fermi energy, $E_F$, at $V=0$~V (blue dashed line) lies in the gap. The upper inset shows the current-voltage characteristics from which the conductance was derived and the lower inset shows an STM topography image taken at 300 K with a bias voltage of -1.0 V. (b) Normalized conductance at 77 K (current set-point respectively 1.5 nA, 2.5 nA, and 4.0 nA at a bias voltage of -4.0 V). The schematic drawing explains the larger gap between the valence band $\Gamma$ point at the STO capping surface and the substrate-LAO interface conduction band. The Fermi energy (blue dashed line) lies now in the conduction band. Band bending is not depicted on this scale. The inset shows the current-voltage characteristics.}
\end{figure}

Figure~\ref{fig:STM}~shows the current-voltage tunneling characteristics of a STO(001)/2LAO/1STO sample and the derived normalized conductivity $(dI/dV)/(I/V)$, which can be interpreted as the sample local density of states (LDOS). Assuming that tunneling occurs both to the surface and to the substrate-LAO interface, these spectroscopic features can be understood on the basis of the band structure calculations. At 300 K, the Fermi energy lies between the valence band at the M-point in the surface and the conduction band at the $\Gamma$-point in the substrate-LAO interface. At 77 K, see Fig.~\ref{fig:STM}b, the valence band seems to have shifted to lower energy. We will elaborate below that this is due to the momentum-resolving capabilities of STS. 

The apparent increase of the STS band gap when cooling down from room temperature to 77 K can be understood from an analysis of measurements of the tunnel current, $I$, versus the tip-sample separation, $z$, at fixed (negative) sample bias. The tunnel current $I \propto e^{-2 \kappa z}$ is measured at negative sample biases in the range from -1 V to -3 V. The inverse decay length, $\kappa$, is substantially larger at room temperature ($\sim$ 80 \%) than at 77 K. The inverse decay length is given by $\kappa=\sqrt{C+k^2_{_{\parallel}}}$, where $C$ only depends on the temperature independent tunnel barrier height and $k_{_{\parallel}}$ is the parallel momentum of the surface electronic structure \cite{Zandvliet}. The relatively large $\kappa$ at room temperature strongly supports the idea that at room temperature tunneling occurs from filled surface electronic states with a nonzero parallel momentum (\textit{i.e.} regions near the M point of the surface Brillouin zone) to empty states of the tip. Accordingly, at 77 K tunneling mainly occurs from filled electronic states near the $\Gamma$ point of the surface Brillouin zone to empty states of the tip, leading to substantial increase of the measured band gap. The latter analysis emphasizes that the electrons that are transferred across the STO/LAO/STO structure upon reconstruction indeed originate from the M point of the surface STO Brillouin zone. 

\begin{figure}%[!h]
     \begin{center}
\includegraphics[scale=0.65]{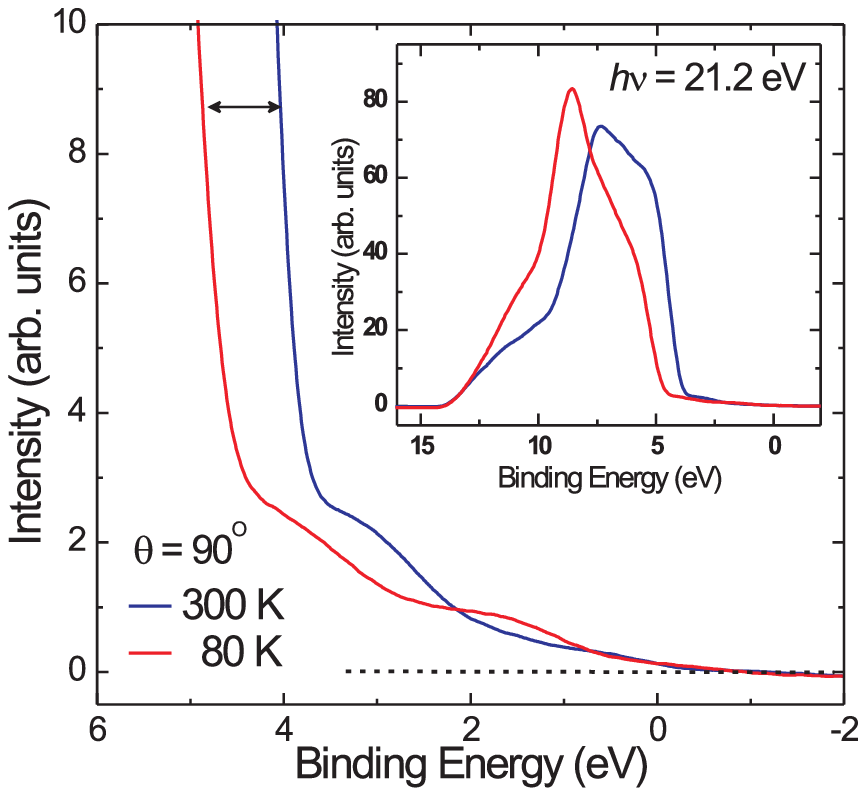}
    \end{center}
\caption{\label{fig:UPS} UPS spectra of a STO(001)/2LAO/1STO sample taken \textit{in situ} after growth, at 300 K and 80 K with the detector at 90$^{\circ}$. The inset shows the zoomed-out full spectra. The data has been normalized to the integrated peak height. The arrows indicate a significant shift that was observed between the spectra. }
\end{figure}

The STS measurements also reveal that, upon lowering the temperature, the Fermi energy is slightly shifted to higher energies. At 77 K, $E_F$ is found to ly in the conduction band. This finding was confirmed by the temperature dependence of the UPS spectra. While the overall spectra (see inset of Fig.~\ref{fig:UPS}) resemble those for bulk STO \cite{UPSWolter}, a clear 1 eV peak shift to higher binding energy is observed for 80 K when compared to 300 K. This can be interpreted as an upward shift of the Fermi energy relative to the valence band maximum in the bulk of STO when lowering the temperature, analogous to the observation by STS.

\begin{figure*}
    \vspace{-10pt} 
\includegraphics[scale=0.635]{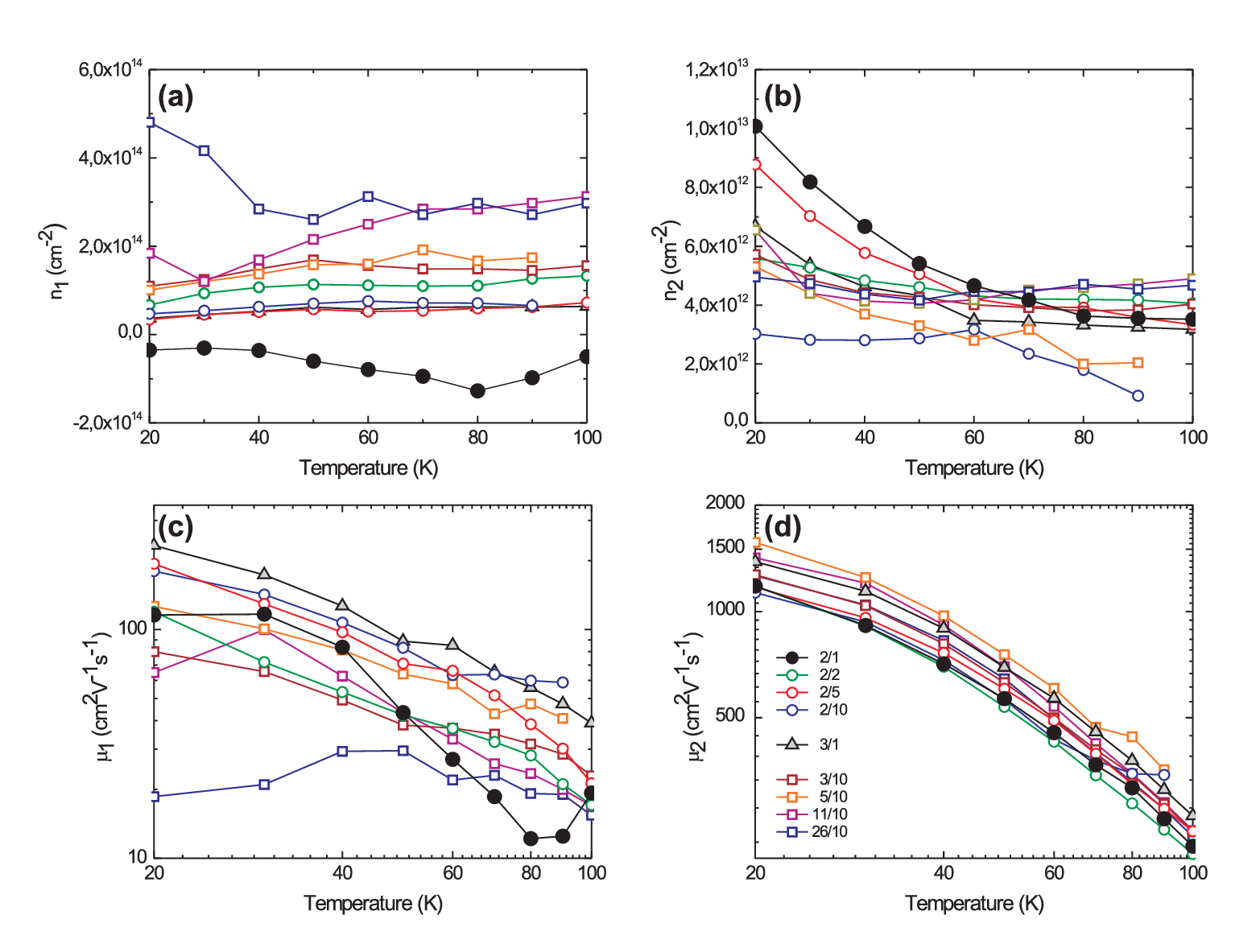}
\caption{\label{fig:S3} Two-band fitting results for different STO(001)/$n$LAO/$m$STO samples. The magnetoresistance and Hall effect field dependence were fitted with a two-band model. The fitting provides two carrier densities (a-b) and two mobilities (c-d) for each $n/m$ sample.}
\end{figure*}

%-----------------------------------------------------------------------------
\section{Two-band fitting results}

For every electronic band $n$, that contributes to conductivity, the induced current $\bm{j_n}$ is given by the electric field $\bm{E_n}$ times the band conductivity $\bm{\sigma_n}$, $\bm{j_n}=\bm{\sigma_n}\bm{E_n}$. The band resistivity $\bm{\rho_n}=\bm{\sigma_n}^{-1}$ is defined as
\begin{equation}
\bm{\rho_n}=\left(
\begin{array}{cc}
\rho_n & -R_{n}H \\
R_{n}H & \rho_n
\end{array}
\right),\nonumber
\end{equation}
where $\rho_n$ is the longitudinal resistance, $R_n$ the transverse Hall coefficient and $H$ the magnetic field.
The \textit{total} resistivity tensor $\bm{\rho}$, defined as 
\begin{equation}
\bm{\rho_n}=\left(
\begin{array}{cc}
\rho & -RH \\
RH & \rho
\end{array}
\right),
\end{equation}
is given by $\bm{\rho}=\bm{\sigma}^{-1}=\left(\sum_n \bm{\sigma_n}\right)^{-1}= \left(\sum_n \bm{\rho_n}^{-1}\right)^{-1}$.
When only two bands contribute to conductivity, it follows that the total longitudinal and Hall resistances can be expressed as
\begin{eqnarray}
\rho=\frac{\rho_0+\rho_{\infty}\mu^2H^2}{1+\mu^2H^2},\label{R}\\
R_H=\frac{R_0+R_{\infty}\mu^2H^2}{1+\mu^2H^2}\label{RH},
\end{eqnarray}
where $R_0 = \left(R_1\sigma_1^2+R_2\sigma_2^2\right)\left(\sigma_1+\sigma_2\right)^{-2}$,
$R_{\infty} = R_1R_2\left(R_1+R_2\right)^{-1}$,
$\mu = \left(R_1+R_2\right)\sigma_1\sigma_2\left(\sigma_1+\sigma_2\right)^{-1}$,
$\rho_0 = \left(\sigma_1+\sigma_2\right)^{-1}$, 
and $\rho_{\infty}=\left(R_1^2\sigma_2^{-1}+R_2^2\sigma_1^{-1}\right)\left(R_1+R_2\right)^{-2}$.

The band conductivities are given by $\sigma_{1,2}=\left|n_{1,2}\right|\mu_{1,2}$ and the band Hall resistivities by $R_{1,2}=\left(n_{1,2}\right)^{-1}$, where $n$ is negative for electrons (negative curvature in the band dispersion relation) and positive for holes (positive band curvature).

Equations (\ref{R}) and (\ref{RH}) were fitted simultaneously to the measured sheet resistance and Hall resistance by means of a least square fitting routine. The resistivity data was symmetrized (average over values at positive and negative fields) in order to exclude a transverse resistance contribution to the longitudinal resistance. The Hall resistivity was anti-symmetrized (difference between values at positive and negative fields) in order to exclude longitudinal components.  All the different samples could be fitted within the experimental error bars. The results for all measured STO(001)/$n$LAO/$m$STO samples are shown in Fig.~5.